\documentclass[
a4paper,
12pt,%
notitlepage,%
centertags]%
{revtex4}

\usepackage{fancyhdr}
 \fancypagestyle{plain} {
\fancyhead{}
}
\usepackage{latexsym}
\usepackage[dvips]{graphicx}
\usepackage{floatflt}
\usepackage[mathscr]{eucal}
\usepackage{graphicx}

\begin{document}
\title{CLASSICAL COSMOLOGICAL TESTS FOR GALAXIES OF THE HUBBLE ULTRA DEEP FIELD}

\author{\firstname{N.~V.}~\surname{Nabokov}}
\email{NabokovNikita@yahoo.com}
\affiliation{Sobolev Astronomical Institute,  St. Petersburg State University, Staryj Peterhoff, St. Petersburg,
198504 Russia}

\author{\firstname{Yu.~V.}~\surname{Baryshev}}
\email{yuba@astro.spbu.ru}
\affiliation{Sobolev Astronomical Institute,  St. Petersburg State University, Staryj Peterhoff, St. Petersburg,
198504 Russia}

\received{November 6, 2007}%
\revised{January 18, 2008}%

\begin{abstract}
Images of the Hubble Ultra Deep Field are analyzed to obtain a
catalog of galaxies for which the angular sizes, surface
brightness, photometric redshifts, and absolute magnitudes are
found. The catalog contains a total of about 4000 galaxies
identified at a high signal-to-noise ratio, which allows the
cosmological relations angular size--redshift and surface
brightness--redshift to be analyzed. The parameters of the
evolution of linear sizes and surface brightness of distant
galaxies in the redshift interval 0.5--6.5 are estimated in terms
of a grid of cosmological models with different density parameters
($\Omega_V; \,\Omega_m $). The distribution of photometric
redshifts of galaxies is analyzed and possible superlarge
inhomogeneities in the radial distribution of galaxies are found
with scale lengths as large as 2000~Mpc.
\end{abstract}

\maketitle

\section{INTRODUCTION}

The program of observational cosmology, which was first formulated
by Hubble and \mbox{Tolman \cite{Tolman:Nabokov_n}} and then
further developed by Sandage
\cite{Sandage:Nabokov_n,Sandage2:Nabokov_n} suggested a number of
cosmological tests---N(m), N$(z)$, $m(z)$, $J(z)$, $\Theta(z)$,
and t$(z)$---using numbers of objects, magnitudes, surface
brightness, sizes, and their ages. These tests, which are also
called classical tests, are based on the comparison of empirical
relations between directly observable quantities with the
theoretical relations between the same quantities as predicted by
different cosmological models.

Modern approach toward the analysis of classical cosmological
tests consists in the simultaneous taking into account both the
parameters of the cosmological model and the evolution of
galaxies. However, so far, no bona fide model of the evolution of
galaxies is available and the development of such a  model remains
the main unsolved problem of modern cosmology.

The evolution of galaxy parameters is usually estimated for the
``standard'' values of cosmological parameters \mbox{$\Omega_{m} =
0.3$} and $\Omega_{V} = 0.7 $ with reference to WMAP data.
However, as Spergel et al.~\cite{Spergel:Nabokov_n} pointed out,
interpretation of observations of the fluctuations of microwave
background includes 15 parameters of the standard model and only
six of them can be estimated independently. Note that the
parameter $\Omega_V$ of vacuum density is not among those six
parameters. Parameter $\Omega_V$ is estimated using a combination
of other observational data, such as the Hubble diagram for type
Ia SNe and the correlation properties of the large-scale
distribution of galaxies.

Note that the values of cosmological parameters estimated using
different methods may differ significantly from the corresponding
``standard'' values. Indeed Clochiatti et
al.~\cite{Clochiatti:Nabokov_n} used SNIa observations in the
``Hight-Z Supernova Search'' program to estimate the following
parameters: \mbox{$\Omega_{m}=0.79 \pm 0.15$} and $\Omega_{V}=1.57
\pm 0.25$. The results obtained in ``The Supernova Cosmology
Project'' \cite{Conley:Nabokov_n} using the light curves of type
Ia supernovas obtained simultaneously in several filters yielded
the estimates \linebreak \mbox{$\Omega_m = 1.26 \pm 0.4$,}
$\Omega_V = 2.20 \pm 0.5$. Thus the estimates of cosmological
parameters $\Omega_{m}$ and $\Omega_{V}$ may vary over a wide
range.

In this paper we perform a quantitative analysis of the dependence
of the parameters of galaxy evolution on the cosmological model
parameters  $\Omega_{m}$ and $\Omega_{V}$. We use results of
observations of the Hubble Ultra Deep Field to obtain the diagrams
``angular size--redshift'' ($\Theta(z)$) and ``surface
brightness--redshift'' $(J(z))$ for galaxies in the redshift
interval 0.5--6.5 and analyze the effect of a change of
cosmological parameters on the estimates of the galaxy evolution
parameters.

In addition, we also analyze the distribution of photometric
redshifts of galaxies---the $dN(z)/dz$ test---for a
magnitude-limited sample. Our method is capable of identifying
large-scale fluctuations of the number of galaxies, which exceed
the Poisson noise level.

\section{OBSERVATIONAL DATA}

Problems involving cosmological tests require reliably
determinable galaxy parameters, i.e., to address such problems,
one must use values of the numerous parameters of photometric
reduction and identification of distant galaxies  and ensure
sufficiently high signal-to-noise ratio. To do this, we compiled a
catalog of galaxies of the Hubble Ultra Deep Field, including
objects with the signal-to-noise ratio higher than 5.

\subsection{Identification of Galaxies}

The data on the Hubble Ultra Deep Field (HUDF) can be found at
{\it \small http://www.stsci.edu/hst/udf}.~ We used the images
taken in four filters adopted from \linebreak {\it \small
http://archive.stsci.edu/prepds/udf/udf\_hlsp.html}, where the
initial reductions have already been performed, allowing the
identification of objects on frames to be immediately addressed.
We used sExtractor \cite{sExtractor_help:Nabokov_n} code to
identify objects. Our input data consisted of a configuration file
and files with the frames taken in four filters (B, V, i, and z).
In the configuration file we set the parameters that we used to
identify objects in the field studied.

We set the~~PIXEL\_SCALE parameter equal to $0.03''$ for the field
considered. The process of identification of objects depends
significantly on parameter DETECT\_THRESH. sExtractor interprets
the signal as a part of the galaxy if the flux level exceed this
parameter.

Our criteria of the identification of objects are based on the following assumptions:
\begin{enumerate}
    \item[1)] all pixels record signals that exceed the given  DETECT\_THRESH;
    \item[2)] pixels form a group (i.e., they are ``crowded'');
    \item [3)]the number of pixels in a given group is greater than the given natural number.
\end{enumerate}
We considered the pixels with flux deviated by more than $3\sigma$
from the mean flux to be parts of the object.

The ``crowdnectess'' of pixels was set by parameter
DETECT\_MINAREA. If a group of pixels has a count above $3\sigma$
and the number of pixels exceeds  DETECT\_MINAREA, the programm
concludes that a galaxy has been detected.

The mean value is estimated by averaging a portion of the field
rather than the entire field. The  given averaging region had the
size of  $100 \times 100$ pixels. The finding of the background
averaging region is an issue of great importance, because use of
too small regions effects an estimate being obtained, and use of
large regions results in a strong effect on the detected objects.

We also additionally smoothed the image. The smoothing procedure
is performed before the detection of galaxies of the field. In
this paper we use Gaussian as the smoothing function. The size of
the smoothing region is $3 \times 3$ pixels. In particular, the $3
\times 3$ region has the form of a square matrix (the axes X and
Y), where each cell stores the corresponding normalized flux (the
Z coordinate). In this way the three-dimensional model of the
smoothing function  is modeled, where X and Y correspond to the
coordinates in pixels and  Z, to the flux.

In cases of the search and separation of objects we used an
additional parameter, which allows to take into account the input
from the bright objects. All detected objects are tested for the
closeness to bright objects. In cases of ``contamination'' by a
bright object we use the following formula of the Moffat profile:

\begin{equation}
\frac{J(r)}{J(0)} = \frac{1}{(1 + k \times r^{2})^\beta},
\end{equation}
\noindent
where $\beta$ is set in the sExtractor configuration file and $J$ is the surface brightness. 
A change in parameter $\beta$ affects significantly the photometry and detection of field objects.

\subsection{Finding the Parameters of Identified Objects}

We found the main parameters of the objects identified.

\subsubsection{Photometric parameters}

\begin{figure}[htbp]
\includegraphics[scale=2.2]{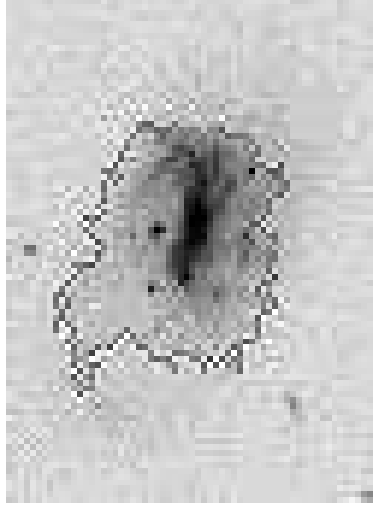}
\caption{Isophote of one of HUDF galaxies.}\label{Isompotal_Area}
\end{figure}

The photometric parameters are characterized by the flux from the
given area of the object. There are several possible methods of
identifying the area the flux is to be computed from. In this
paper we use the ``isophotal'' approximation set in the sExtractor
configuration file. In this method the contours of the area are
found from the count level that depends on the average flux in the
given area. Figure~\ref{Isompotal_Area} shows the
isophote of one of the galaxies in the field studied. The
background is subtracted from the flux gathered from the given
area and the magnitude of the detected object is computed.

While compiling the catalog of objects we tried to use different
methods of flux computation and found no significant differences
between the results of the photometry of galaxies (depending on
the method). This may be due to the fact that most of the galaxies
in the field studied have small sizes and irregular structure.

The instrumental flux is related to magnitudes via the following formula:

\begin{equation}
m = 2.5 \times \lg(F) + m_{zp},
\end{equation}
where $F$ is the flux in instrumental units and $m_{zp}$, the
average magnitude of background for each filter. We also find the
maximum surface brightness of the object, which is available via
parameter MU\_MAX in mag$^{''}$

The input catalog of objects contains the following photometric information:

\begin{itemize}
    \item The instrumental flux with its error;
    \item The apparent magnitude of the object with its error;
    \item The maximum surface brightness of the galaxy;
    \item The effective radii corresponding to  25\%, 50\%, and 75\% of the flux of the entire galaxy.
\end{itemize}

\subsubsection{Astronomical parameters of galaxies}

We use the barycentric coordinates of the objects computed by  sExtractor in accordance with the following
formula:

\begin{equation}
X = \bar{x}=\frac {\sum{I_{i}\times x_{i}}} {\sum{I_{i}}},
\end{equation}
\noindent
where $I_{i}$ are the moments equal to the galaxy flux in each pixel.

The coordinates of the objects are computed both in the equatorial coordinate system for the epoch of 2000.0
and in the relative (Cartesian) coordinate system. The relative coordinate system is determined by the coordinates
of galaxies in terms of image pixels.

\subsubsection{Geometric parameters}

The geometric parameters describe the sizes and appearance of the
objects. The ellipticity of galaxies is characterized by the
semiminor and semimajor axes ({\it a} and {\it b}, respectively)
and by position angle $\Theta_{se}$. The semiminor and semimajor
axes are computed using second-order moments.

The formulas for the second-order moments have the following form:

\begin{equation}
\overline{x^{2}} = \frac {\sum{I_{i}\times x_{i}^2}} {\sum{I_{i}}}
- \bar{x}^2
\end{equation}

\begin{equation}
\overline{y^{2}} = \frac {\sum{I_{i}\times y_{i}^2}} {\sum{I_{i}}}
- \bar{y}^2
\end{equation}

\begin{equation}
\overline{xy} = \frac {\sum{I_{i}\times x_{i} \times y_{i}}}
{\sum{I_{i}}} - \bar{x} \times \bar{y}.
\end{equation}

The semiaxes can then be computed by the following formulas:

\begin{equation}
a^{2} = \frac {\overline{x^2} + \overline{y^2}} {2} +
\sqrt{\left(\frac{\overline{x^2} - \overline{y^2}}{2} \right) +
\overline{xy}^2}
\end{equation}

\begin{equation}
b^{2} = \frac {\overline{x^2} - \overline{y^2}} {2} +
\sqrt{\left(\frac{\overline{x^2} - \overline{y^2}}{2} \right) +
\overline{xy}^2}.
\end{equation}

The position angle $\Theta_{se}$ is counted from the North direction for the epoch of 2000.0.

The oblongness of the object is characterized by its ellipticity ($1 - \frac{b}{a}$) or elongation ($\frac{a}{b}$).

We also compute the area of the object at the level of the  detection threshold DETECT\_THRESH.

\subsubsection{Sizes of galaxies}

Cosmological tests usually employ the effective sizes of galaxies
at half the level of flux profile---the so-called  FWHM (Full
Width at Half-Maximum). We described the flux profile by a
Gaussian.

\subsection{Identification of Objects}

After setting the parameters in the  sExtractor configuration file
we created a file used to identify objects in the entire field. We
obtained it by coadding the  FITS files (frames of the field
studied taken in four filters) using MIDAS package. We did it
because objects appear greater in the  i and z filters compared to
their sizes in the  B and V filters. To compile the preliminary
catalog in each filter, we use the file of object detection and
the file with the image. As a result, we found more than 4300
objects.

We then produced a sample of objects by imposing the following
constraints. First, the signal-to-noise ratio (S/N) must be no
lower than 5. We excluded from the sample all objects with S/N
smaller than~5. We further excluded all objects with no measured
flux in at least one of the four filters. As a result, we obtained
a catalog containing  4125 galaxies. 

\subsection{Finding the Redshifts}

After identifying the objects and compiling the preliminary
catalog in four filters we found the photometric redshifts z and
absolute magnitudes of galaxies.

The photometric redshifts of galaxies are inferred from their
magnitudes in different filters. The efficiency of this method is
based on identifying the photometric data points to portions of
the continuum spectrum of the galaxy. The accuracy of photometric
estimates is lower than that of spectroscopic estimates and
depends on the set of filters employed and the accuracy of
photometric data. However, photometric redshifts are quite
suitable for many cosmological and extragalactic problems. This
method of redshift finding has been playing important part in
observational cosmology.

Although finding of photometric redshifts requires no
spectroscopic data, we must, however, trust the template galaxy
spectra used for comparison.

We use HyperZ code \cite{HyperZ_help:Nabokov_n} to compute the photometric redshifts.

The input parameters of HyperZ code include:
\begin{itemize}
  \item Apparent magnitudes of objects in the filters B, V,~ i, and z and their errors;
  \item Template spectral energy distributions (SED);
  \item The reddening law for the objects;
  \item Cosmological parameters.
\end{itemize}

\subsubsection{The method of redshift finding}

The procedure of the finding of photometric redshifts is based on
the comparison of the observed energy distribution in the spectrum
of the galaxy with the template spectral energy distribution. The
observed distribution (inferred from photometric data) is compared
to different template galaxy spectra using the same photometric
system.

We find the photometric redshift $z$ for the given object from the
template spectrum that fits best the observed spectrum. The
fitting procedure is based on minimization of  $\chi^2$. We
compare the observed and template distributions by the following
formula:
\begin{equation}
\chi^{2}(z) = \sum_{i=1}^N \Big[\frac{F_{obs,i} - b\times
F_{temp,i}(z)}{\sigma_{i}}\Big]^2,
\end{equation}
where $F_{obs,i}$ is the observed flux; $F_{temp,}$, the template
flux; $\sigma_{i}$, the variation of the flux in the given filter,
and $b$, the normalization constant. We thus find the redshift by
minimizing $\chi^{2}$.

HyperZ code includes both observed \cite{Coleman:Nabokov_n} and
synthetic template spectra. Template spectra can be replaced and
are stored in the configuration file.

As the synthetic models of galaxy spectra we use  GISSEL98 datasets (Galaxy Isochrone Synthesis Evolution
Library), which feature spectra from a wide interval of energies: from UV to far IR with the possibility of
evolution to large $z$.

Note that we computed the photometric redshifts using template redshifts for galaxies of the E/S0, Sa, Sb, Sc,
Sd, Im, and Burst types.

\subsubsection{Various corrections and reductions}

\textit{Metallicity.} HyperZ code can take into account the
evolution of galaxy metallicity. To do this, special template
galaxy spectra must be indicated that take this factor into
account. We found the photometric  $z$ using template spectra with
and without the accounting for metallicity evolution. The results
obtained led us to conclude that metallicity evolution has no
effect on the inferred photometric redshifts. That is why we did
not take metallicity evolution into account while finding $z$.

\textit{Reddening law.} Recent studies of galaxies at large $z$
demonstrated the importance of the accounting for the
``reddening'' when
estimating the redshifts because of the effect of Galactic dust in the object studied. 
The Calzetti law \cite{Calzetti:Nabokov_n} describes fairly well the extinction for objects at large redshifts
and we adopted it in this paper. Calzetti et al.~\cite{Calzetti:Nabokov_n} describe the empirical results
for the group of young galaxies, which we  used for finding $z$:\\

\begin{eqnarray}
k(\lambda) = 2.659\left(-2.156+\frac{1.509}{\lambda}-\frac{0.198}{\lambda^2}+ \frac{0.011}{\lambda^3}\right)+ R_V,
0.12\mu m \leq \lambda \leq 0.63\mu m \quad \quad \nonumber\\
k(\lambda) = 2.659\left(-1.857+\frac{1.040}{\lambda}\right) +R_V, 0.63\mu m \leq \lambda \leq 2.20\mu m\quad \quad,
\end{eqnarray}

\noindent where $R_V = 4.05$---is the total extinction in the
V-filter.

This law corresponds to central star-forming regions and it can
therefore be applied to galaxies at large redshifts. Note that
extinction is important in the UV part of the spectrum. As a
result, extinction must show up appreciably for galaxies with
\mbox{$z \geq 3$} in the optical part of the spectrum where the UV
portion has been shifted.

\textit{Further corrections.}
The radiation of distant galaxies is also  ``distorted'' by extinction in HI regions located along the line
of sight, which shows up in the part of the spectrum at wavelengths shorter than
\mbox{Ly$\alpha(\lambda = 1216$\AA).} HyperZ code allows for this effect and applies corrections in accordance
with the law suggested by Oke and Korycansy~\cite{Oke_Korycansky:Nabokov_n}.

\subsubsection{Parameters used to find the redshifts}

To find the photometric redshifts,  HyperZ code needs a
configuration file with the following data:
\begin{itemize}
    \item Different cosmological models with \mbox{$H_0 = 72$ km/s $\times$ Mpc$^{-1}$} for computing
     absolute magnitudes $M$ in the B filter;
    \item The magnitudes from the input catalog of galaxies in various filters;
    \item The transmission parameters of the  B, V, i, and z-band filters in accordance with HST data;
    \item The redshift step,  $\Delta z = 0.1$.
\end{itemize}

\subsection{Finding the Surface Brightness Profiles of the Galaxies}

The surface brightness profiles of galaxies are computed via iterations by the following formula:
\begin{equation}
\chi^{2} = \sum_{x=1}^{nx} \sum_{y=1}^{ny}
\Big[\frac{(flux_{x,y}-model_{x,y})^{2}}{\sigma_{x,y}^{2}}\Big],
\end{equation}
where $flux_{x,y}$ and $model_{x,y}$ correspond to the observed
and model flux, respectively and $\sigma$ to the background
weight. We minimize $\chi^{2}$ over the entire image (the
coordinates $x$ and $y$) and use GALFIT for
computations~\cite{GALFIT:Nabokov_n}.

We fit the surface-brightness distribution by a Gaussian. It would make no sense to use a more complex model
for the surface-brightness distribution, because most of the galaxies in the field studied do not fit
standard classification and have irregular appearance.

We adopt the model surface-brightness profile in the following form:
\begin{equation}
J_{r}=J_{0}\times \exp\Big(\frac{-r^{2}}{2\sigma^{2}}\Big),
\end{equation}
where $J_{0}$ is the central surface brightness; $r$, the
galactocentric distance, and $ \sigma = FWHM_{sb} / 2.345$. (here
$FWHM_{sb}$ is the full width of the Gaussian at half maximum). We
compute  $J_{0}$ by the following formula:
\begin{equation}
J_{0}=\frac{F_{tot}\times R(c)}{2\pi\sigma^{2}q},
\end{equation}
where $F_{tot}$ is the total flux from the object studied and  $q$, one of the input approximation parameters
of GALFIT code,
\begin{equation}
R(c)=\frac{\pi(c+2)}{4\beta(1/(c+2),1 + 1/(c+2))};
\end{equation}
$\beta$ is the Beta function, and $c$ is the galaxy ellipticity parameter.\\
We set the following input parameters for finding of the
surface-brightness profiles:
\begin{itemize}
  \item The size of the field containing the object (a FITS file);
  \item The size of the field to be fitted;
  \item The background level in magnitudes;
  \item The $\sigma$ value of the background for each pixel in the given region of the field
  (the weight  image, a FITS file);
  \item The pixel scale in arcseconds.
\end{itemize}\textit{}\\

\begin{figure*}[htbp]
\includegraphics[scale=1.7]{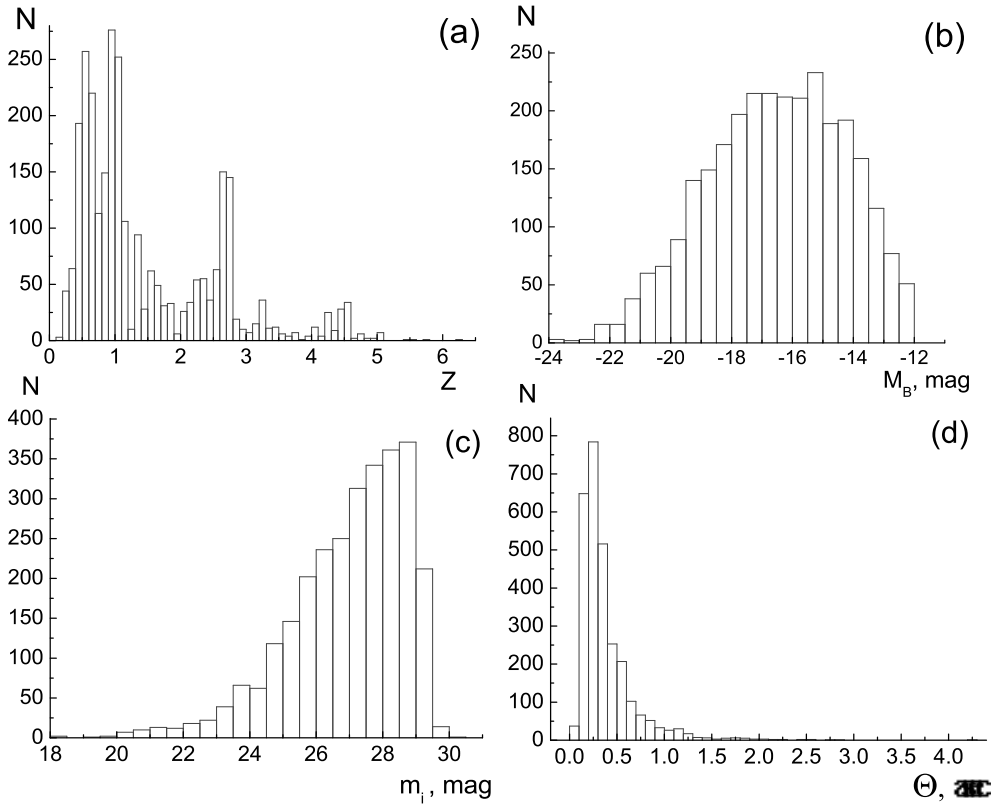}
\caption{The distribution of the observed
quantities of the galaxies of the catalog for the
$C_{i}\_bviz\_mod3$ model, where (a) is the distribution of
photometric redshifts; (b)---the distribution of B-band absolute
magnitudes; (c)---the distribution of i-band apparent magnitudes,
and (d)---the distribution of angular sizes.}
\label{Catalog_Distribution.EPS:Nabokov_n}
\end{figure*}

The initial data (the zero approximation of the theoretical model of surface-brightness distribution)
include:
\begin{itemize}
  \item The coordinates of the center of the object;
  \item The total flux from the entire object in magnitudes;
  \item The $FWHM$ of the surface-brightness profile of the object;
  \item The axial ratio b/a of the galaxy;
  \item The position angle of the galaxy.
\end{itemize}
We computed the surface-brightness profile only for galaxies with
absolute magnitudes  $M$ ranging from -20 to -18 and with the
99-100\% probability of photometric redshifts ($z_{phot}$) (it is
one of the parameters in the general catalog of objects). After
selecting objects from the catalog in accordance with the above
criterion we studied  the objects with $FWHM_{flux} > 10$ pixels.
We imposed this additional constraint on the sample in order to
improve the convergence of the method that we use to find the
surface-brightness profile.

As a result, we obtained a catalog of parameters for finding the
surface-brightness profiles in the
Gaussian approximation for each object of the sample considered.\\

\section{ANALYSIS OF THE GALAXY SAMPLE}

\subsection{Constructing the Main Catalog}

We already mentioned above that after reducing the HUDF we
constructed the general catalog of the objects. However, we then
modified this catalog and drew various samples from it. To
systematize the data, here we review the samples and catalogs.

\begin{figure*}[htbp]
\includegraphics[scale=1.5]{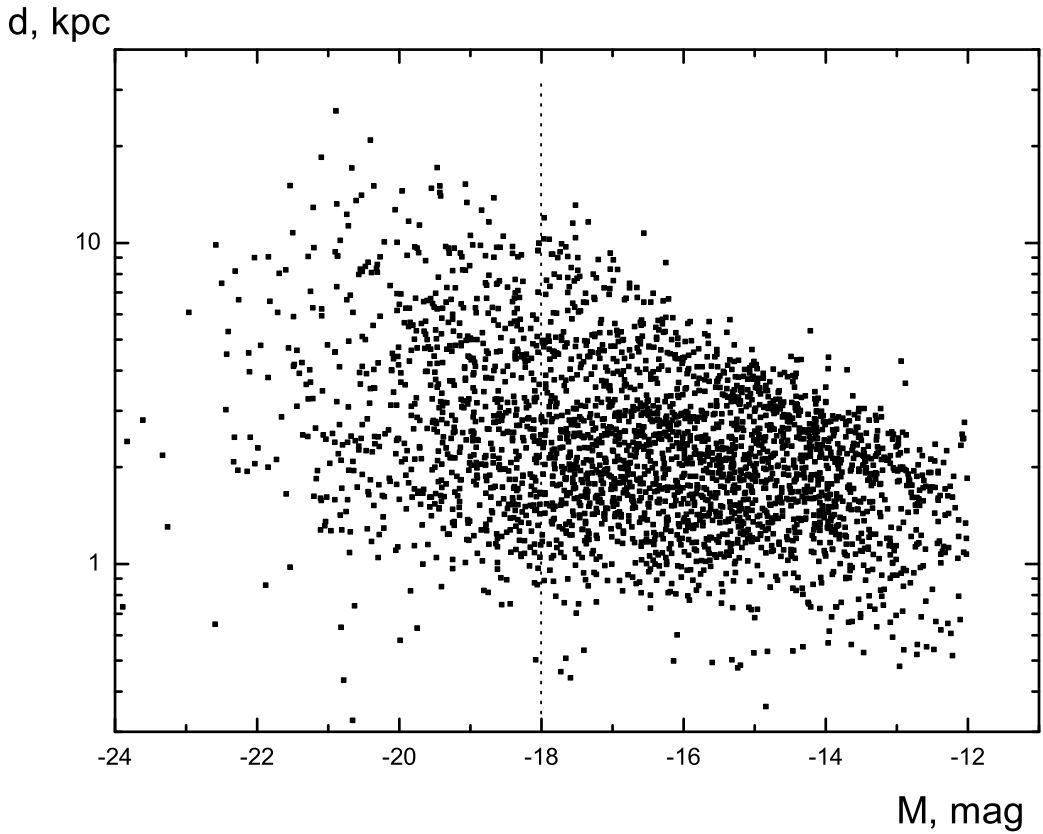}
\caption{The ``absolute magnitude--linear
size'' relation for all galaxies of the sample $C_{i}\_bviz\_mod3$
with the parameters of standard cosmological
model.}\label{M_abs_d_distrib_full_23_06 2007:Nabokov_n}
\end{figure*}

After the objects were identified by sExtractor, we constructed
the preliminary catalog of data, $C_{i}\_bviz$. It contains the
principal parameters (the coordinates, sizes, ellipticity, fluxes,
etc.) for each object in the four filters B, V, z, i.

To find the photometric redshifts, we compiled the input catalog
of data containing the magnitudes of the galaxies in the four
filters and their color indices. We added the input data including
$z_{ph}$ (photometric redshifts), $M$ (absolute magnitudes), and
Galaxy Types into the the general data catalogs
$C_{i}\_bviz\_mod1$, $C_{i}\_bviz\_mod2$, $C_{i}\_bviz\_mod3$, and
$C_{i}\_bviz\_mod4$ corresponding to the cosmological models,
\mbox{$\Omega_V=0.0$} $\Omega_m=0.0$, $\Omega_V=0.0$
\mbox{$\Omega_m=1.0$}, $\Omega_V=0.7$ $\Omega_m=0.3$, and
$\Omega_V=1.0$ \mbox{$\Omega_m=0.0$.} We then subscripted the name
of the catalog depending on the absolute magnitude of objects
selected from the main catalog: $M18\_20$ (galaxies with absolute
magnitudes $M$ in the interval from --20 to --18) or $M20\_22$.

We the excluded from the resulting catalog of objects the galaxies
with incorrectly computed flux in at least one of the filters (in
these cases sExtractor outputs the the flux value equal to 99
magnitudes) and the probability of finding the photometric
redshift $z_{ph} < 90\%$.

\subsection{Distribution of the Observed Parameters of the HUDF Galaxies}

In Fig.~\ref{Catalog_Distribution.EPS:Nabokov_n} we represent the
distribution of the observed parameters of the objects of the
i-band catalog. The histograms of  $\Theta$, $m$, $M$, and $z$ are
computed for the galaxy sample of the $C_{i}\_bviz\_mod3$ catalog,
which contains a total of 4125 objects.

It is evident from Fig.~\ref{Catalog_Distribution.EPS:Nabokov_n}
that the distribution of photometric redshifts is significantly
inhomogeneous~(a);  the distribution of absolute magnitudes  $M$
lies in the interval from --24 to  --12 (b); the number of
galaxies increases with decreasing flux down to  $29^m$, which
corresponds to the completeness limit of the catalog  (c); the
characteristic angular size $\Theta$ is equal to $0.3''$  (d).

\subsection{Constraints on Absolute Magnitudes}

Figure~\ref{M_abs_d_distrib_full_23_06 2007:Nabokov_n} shows the
``absolute magnitude---linear size'' diagram for galaxies. We use
the $FWHM_{flux}$ quantities, which correspond to the halfwidth of
the flux profile, as the sizes of the corresponding galaxies.

The increase of linear size becomes slower for bright galaxies and
therefore, to reduce the systematic displacement of the data
points on the  $\Theta(z)$ diagram, constraints are to be imposed
on the absolute magnitudes of galaxies. The distribution of linear
sizes of faint galaxies is uniform and therefore we exclude them
in order to reduce the systematic errors. The vertical line in
Fig.\ref{M_abs_d_distrib_full_23_06 2007:Nabokov_n} corresponds to
the limiting absolute magnitude of galaxies included into the
sample, which contains galaxies with absolute magnitudes
\mbox{brighter than --18.}
We compute the absolute magnitudes using formula (\ref{62:Nabokov_n}), which already incorporates a fixed
cosmological model and spectral energy distributions of galaxies. Therefore to estimate the evolution parameter
k, we use the subsamples of galaxies limited by absolute magnitude in accordance with each particular
cosmological model.

\subsection{Division into Subsamples}

We found the parameter of the evolution of linear sizes of
galaxies as a function of their type and subdivided the sample
into the subsamples of spiral and elliptical galaxies. HyperZ code
determine the galaxy type by the form of its spectrum while
computing the photometric redshifts. We selected the brightest
galaxies of the catalog for various intervals of absolute
magnitudes. We computed the absolute magnitude $M$ separately for
each cosmological model. Then we subdivided the $C_{i}\_bviz$
catalog into four subsamples (in accordance to each cosmological
model), which we then subdivided into subcatalogs: \mbox{$-20 \leq
M \leq
-18$;} \mbox{ $-22 \leq M \leq -20$}.

\section{THEORETICAL $\Theta(z)$ AND  $J(z)$ RELATIONS}

Many papers have been dedicated to the discussion of observational
tests of cosmological models (see, e.g., the reviews by
Sandage~\cite{Sandage:Nabokov_n, Sandage2:Nabokov_n}). Modern data
are indicative of the need for the use of cosmological models
including both dark matter and dark energy. In this section we
give the general theoretical relations used to analyze the
classical cosmological tests.

\subsection{The ``Metric Distance--Redshift'' Relation}

The density parameter in the Friedmann cosmological models including cold dark matter (CDM) and dark energy
is equal to the sum
\begin{equation}
  \Omega = \Omega_m + \Omega_{V}\,,
\end{equation}
where $\Omega_m = \rho_m/\rho_{crit}$ is the parameter of the density of cold matter and
$\Omega_{V} = \rho_{V}/\rho_{crit}$ is the parameter of the density of dark energy, for which $p_{V}=w\rho_{V}c^2$,
and \mbox{$w \leq 0$}. In the particular case $w=-1$ dark energy corresponds to cosmological vacuum and
Einstein's cosmological constant.

In the general case the Friedmann equation has the following form:
\begin{equation}
\label{f1:Nabokov_n}
 \Omega = 1 - \Omega_k\,,
\end{equation}
or
\begin{equation}
\label{f1-2:Nabokov_n}
 H^2 - \frac{8\pi G}{3} \rho = -\frac{kc^2}{S^2}\,,
\end{equation}
where~~ $\Omega = \varrho / \varrho_{crit}$ is determined by the
total density~~ $\varrho=\varrho_m + \varrho_{V}$; the critical
density is \linebreak \mbox{$\varrho_{crit}=3H^2/8\pi G$}, and the
curvature density parameter is $\Omega_k = kc^2/S^2H^2$. The
Hubble parameter $H =\dot{S}/S $ and the scale factor $S(t)$ are
set by the Robertson--Walker four-interval, which has the
following form:
\begin{eqnarray}
 ds^2 = c^2dt^2 - S^2(t)d\chi^2 - S^2(t) I_k^2 (\chi) (d\theta^2 + \sin^2 \theta d\phi^2)\,,
\label{rw1:Nabokov_n}
\end{eqnarray}
where $I_k(\chi) = \sin(\chi),~\chi,~\sinh(\chi)$ with $k$ = +1, 0,
and --1, respectively.

The proper metric distance $r$ from the observer to the galaxy with dimensionless comoving coordinate $\chi$
in metrics (\ref{rw1:Nabokov_n}) is given by the following formula:
\begin{equation}
  r(t, \chi) = S(t)\chi ~.
\label{rprop:Nabokov_n}
\end{equation}
Note that the proper metric distance $r\,\,$ (measured inside the
three-dimensional hypersphere) and scale factor $S\,$ have the
dimensions of length: \linebreak $[r]=[S]=[cm]$ .

To describe the ``angular size--redshift'' relation, observational
cosmology uses the ``external''$\,$ metric distance $l$
\begin{equation}
\label{rext:Nabokov_n}
  l(t, \mu)=S(t)\mu ~,
\end{equation}
where dimensionless comoving distance $\mu$ appears explicitly in the four-interval
\begin{eqnarray}
\label{rw2:Nabokov_n}
  ds^{2} = c^{2}dt^{2} - S^{2}(t)
  \frac{d\mu^{2}}{1-k\mu^2} - S(t)^{2} \mu^{2}
 (d\theta^{2}+\sin^{2}\theta d\phi^{2}).
\end{eqnarray}
According to formula (\ref{rw2:Nabokov_n}), distance $l=S\mu$ is measured in the ambient four-dimensional space.
The relation between $\chi$ and $\mu\,\,$ ($\chi = I_k^{-1}(\mu), ~~
\mu=I_k(\chi)$) can be used to write the relations between metric distances (\ref{rprop:Nabokov_n}) and
(\ref{rext:Nabokov_n}) in the following form
\begin{equation}
\label{mdrel:Nabokov_n}
  r = S(t) I^{-1}_k(l/S)\,, \qquad l=S(t)I_k(r/S)\,,
\end{equation}
where  $I^{-1}_k$ is the inverse function to $I_k$. At $k=0$ we have $r=l$.

The general formula for metric distance in the Friedmann model has the following form:
\begin{equation}\label{r-z:Nabokov_n}
  r(z)= \frac{c}{H_0}\int_0^z\frac{dz'}{h(z')}\,,
\end{equation}
where $h(z)$ can be derived from the Friedmann equation in the following form:
\begin{equation}\label{h-z:Nabokov_n}
  h(z)=\sqrt{\tilde{\rho}(z)\Omega_0 + (1-\Omega_0)(1+z)^2}\,,
\end{equation}
where $\Omega_0=\rho^0_{\rm tot}/\rho^0_{\rm crit}$ is the density parameter at the present epoch and
$\tilde{\rho}(z)=\rho/\rho_0$ is the normalized total density of all components.

In case of the two-liquid dust + vacuum model (without
interaction) the proper metric distance is given by the following
formula:
\begin{eqnarray}\label{r-z-v+m:Nabokov_n}
  r(z)= \frac{c}{H_0}\times\quad\quad\quad\quad\quad\quad\quad\quad\quad\quad\quad\quad\quad\nonumber\\
  \int_0^z\frac{dz'}{\sqrt{(\Omega_{V}^0 + \Omega_m^0(1+z')^3-\Omega_k^0(1+z')^2)} }\,,
\end{eqnarray}
and the external metric distance is equal to (for  \mbox{$k=-1$}):
\begin{eqnarray}
\label{l-z-v+m-1:Nabokov_n}
 l(z)=\frac{c}{H_{0}}
 \frac{1}{(-\Omega_{k}^0)^{1/2}}\times \quad\quad\quad\quad\quad\quad\quad\quad\quad\nonumber\\
 \sinh \left(
 \int^{1}_{\frac{1}{1+z}}\frac{(- \Omega_{k}^0)^{1/2}dy}
{y\sqrt{( \Omega_{m}^0/y
 - \Omega_{k}^0 + \Omega_{V}^0y^{2})}}\right).
\end{eqnarray}
In the   $k=0$ case :
\begin{equation}
\label{l-z-m+v-0:Nabokov_n}
 l(z)=\frac{c}{H_{0}} \int^{1}_{\frac{1}{1+z}}\frac{dy}
 {y\sqrt{(\Omega_{m}^0/y+ \Omega_{V}^0y^{2})}}.
\end{equation}
In the  $k=+1$ case:
\begin{eqnarray}\label{l-z-m+v+1:Nabokov_n}
 l(z)=\frac{c}{H_{0}} \frac{1}{(\Omega_{k}^0)^{1/2}} \times\quad\quad\quad\quad\quad\quad\quad\quad\quad\quad \nonumber\\
 \sin\left(\int^{1}_{\frac{1}{1+z}}
 \frac{(\Omega_{k}^0)^{1/2}dy} { y (\Omega_{m}^0 / y - \Omega_{k}^0 +
 \Omega_{V}^0y^{2})^{1/2}}\right).
\end{eqnarray}

\subsection{The ``Angular Size--Redshift'' Relation}

The relation between metric distance $l(z)$ and angular size $\Theta$ has the following form:
\begin{equation}
\label{eq21:Nabokov_n}
  \Theta(z) =
  d\left(\frac{1+z}{l(z)}\right)=
  \frac{d}{R_{H\displaystyle_0}}\left(\frac{1+z}{x(z)}\right),
\end{equation}
where $d$ is the fixed size of the galaxy; {\small
\mbox{$x(z)=l(z)/R_{H_0}$}} and $R_{H\displaystyle_0} = c / H_0$,
$H_0 = 72$\,\, (km/s)/Mpc. If $\Theta$ is in arcsec and $d$ in kpc
then formula (\ref{eq21:Nabokov_n}) acquires the following form:
\begin{equation}
\label{eq22:Nabokov_n}
  \Theta(z) =
  0.0481\times d\times\left(\frac{x(z)}{1+z}\right)^{-1}\,.
\end{equation}

\begin{figure*}[htbp]
\includegraphics[scale=1.3]{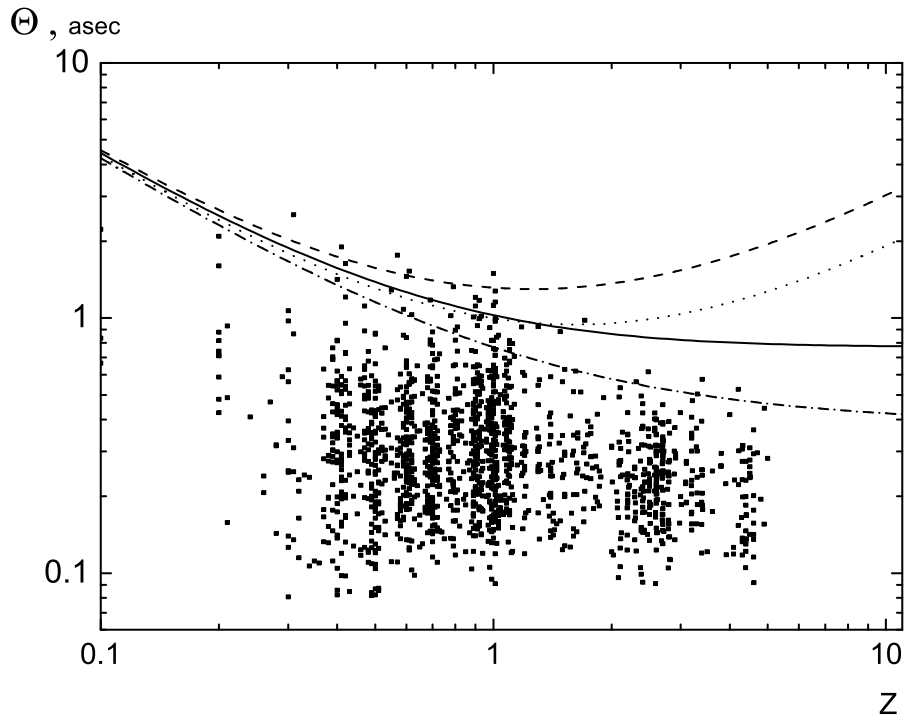}
\caption{Comparison of theoretical models
with observational data for galaxies in the interval of absolute
magnitudes $M$ from  --18 to --22 of the
$C_{i}\_bviz\_el\_M18\_22$ sample. The dashed line corresponds to
the theoretical model~ $\Omega_m = 1.0$ and $\Omega_V = 0.0$; the
solid line---to the model $\Omega_m = 0.0\, \mbox{and}\,\Omega_V =
0.0$; the dotted line---to $\Omega_m= 0.3\, \mbox{and}\,\Omega_V =
0.7$, and the dashed-and-dotted line---to $\Omega_m =0.0\,
\mbox{and}\,\Omega_V = 1.0$.}\label{All_Modeles:Nabokov_n}
\end{figure*}

For different combinations of $\Omega_{m}$ and $\Omega_{V}$ metric
distance $l(z)$ can be computed by formulae
\mbox{(\ref{l-z-v+m-1:Nabokov_n} -- \ref{l-z-m+v+1:Nabokov_n})}.

\subsection{Absolute Magnitudes}
To consider fixed intervals of galaxy luminosities, one must compute the absolute magnitudes of galaxies.
In the Friedmann models the absolute magnitudes are computed using bolometric distance determined as
\mbox{$l_{bol}=l(z)\times (1+z)$}. The absolute magnitude of the galaxy, $\:M_{j}$, in filter $j$ can be computed
from the following general formula:
\begin{equation} \label{62:Nabokov_n}
M_{j} =  m_{j} - 5 \log \{l(z)(1+z)\} - 25 -\Delta M_j \,,
\end{equation}
where distance $l(z)$ is in Mpc and {\small \mbox{$\Delta M_j =
A_j + K_j + E_j$}} is the correction to the absolute magnitude for
extinction, redshift, and luminosity evolution.

\subsection{The ``Surface Brightness--Redshift'' Test}

The ``surface brightness--redshift'' test is a critical one,
because, as Tolman \cite{Tolman:Nabokov_n} was the first to point
out that the $J(z)$ relation is universal and the same for all
Friedmann models.

It follows from the definition of the surface brightness of an object that:
\begin{equation}
    J_{bol} = \frac{F_{bol}}{\Theta^2}= \frac{J_0}{(1 + z)^4},
\end{equation}
where $J_{bol}$ is the bolometric surface brightness; $F_{bol}$,
the bolometric flux, and $J_0$, the surface brightness of the
galaxy at z=0. In case of i-band we are dealing with the
brightness observed in the given wavelength interval rather than
with the bolometric brightness. Therefore below by $J_{bol}$ we
mean $J_{obs}$ (observed in the the i filter). We then have:
\begin{eqnarray}
\mu = \mu_0 + 2.5\log(1 + z)^4 + K_i(z) + E_i(z) = \mu_0 + 2.5\log(1 +
z)^n,
\end{eqnarray}
where $\mu$ is the surface brightness measured in units
mag$^{''}$; $K_i(z)$, the K-correction to the i-band surface
brightness; $E_i(z)$, the evolutionary correction to the i-band
surface brightness; \mbox{$n = 4 + p =$} \mbox{$4 + e_k + e_e$},
and $p$, the combined parameter of the surface-brightness
evolution.

\section{EVOLUTION OF SIZES AND SURFACE BRIGHTNESS}

\subsection{Parameter of the Evolution of Linear Sizes of Galaxies}

Figure~\ref{All_Modeles:Nabokov_n} compares the observational data for galaxies of the $C_{i}\_bviz\_el$ sample
(here  ``el'' indicates that the sample consists of elliptical galaxies) with theoretical models in the
interval of absolute magnitudes  $M$ from --22 to --18.

\begin{figure*}[htbp]
\includegraphics[scale=1.6]{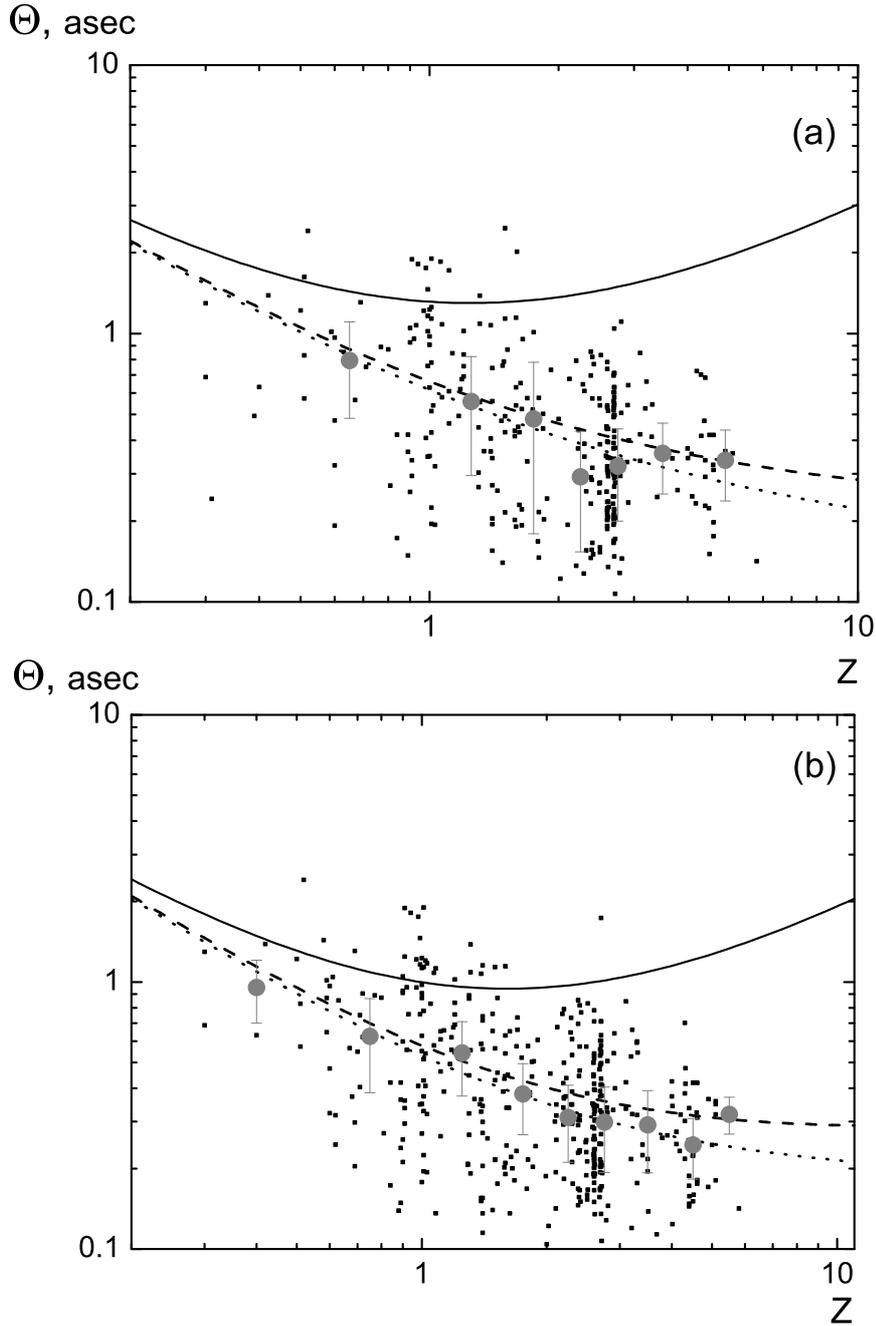}
\caption{Finding the parameters of evolution
for different models of galaxies with $M$ from --20 to --18 for
the $C_{i}\_bviz\_mod3\_M18\_20$ (a) and
$C_{i}\_bviz\_mod4\_M18\_20$ (b) samples. The solid line in plots
(a) and (b) corresponds to the models with $\Omega_m = 1.0,\,
\Omega_V = 0.0$ and $\Omega_m = 0.3,\,\Omega_V= 0.7$,
respectively; the dashed line shows the fit for the entire sample,
and the dotted line, the fit for median points.
}\label{Fittings_a_b_M_-18_-20:Nabokov_n}
\end{figure*}

It is evident from Fig. \ref{All_Modeles:Nabokov_n} that neither model passes through the median values for the
given sample. This fact is usually explained by the evolution of linear sizes of galaxies and function $f(z)$ is
introduced:
\begin{equation}
  \Theta(z)_{obs} = f(z)\times \Theta(z)_{theor},
\end{equation}
where $\Theta(z)_{obs}$ are the observed angular sizes and
$\Theta(z)_{theor}$ are the theoretical angular sizes computed in
terms of the model studied. The function usually has the form
$f(z) = (1 + z)^{k}$, where k is the parameter of evolution.

We compute the parameter of the evolution of galaxy sizes in
accordance with formula (\ref{eq21:Nabokov_n}). In Fig.~5 we
present several plots for models with
$\Omega_m$=1.0,\,$\Omega_V$=0.0 and $\Omega_m$=0.3,
\,$\Omega_V$=0.7. We determine parameter k by applying the method
of least squares both to the values for the entire sample and to
median points. We perform our computations using mathematical
package Microcal Origin 7.0. Note that the {\it median} is a
statistically more stable parameter than the  {\it mean}.
As a result, we find the parameter of the evolution of galaxy
sizes for four models both for the sample with $M$ from --20 to
--18 and for objects with $M$ from  --22 to --20.
Table~\ref{FWHM_flux_values:Nabokov_n} lists the results obtained
for angular sizes.

\begin{figure*}[htbp]
\includegraphics[scale=1.0]{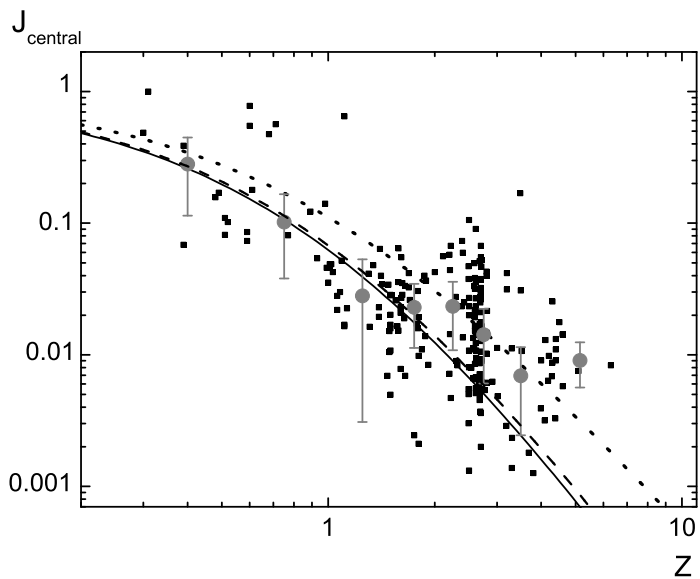}
\caption{Evolution of surface brightness as a
function of  $z$ for the $C_{i}\_bviz\_mod3\_M20\_22$ sample. The
dashed line corresponds to averaging over all points of the
sample; the dotted line---to averaging over the median points; the
solid line shows the theoretical evolution of surface
brightness,and the vertical bars show the errors of the median
values.}\label{Inorm_M-22_-20_fitting:Nabokov_n}
\end{figure*}

Note that we set parameter $d$ equal to 12~kpc for the sample of
objects with $M$ from --22 to --20 (compared to $d$=8~kpc adopted
for galaxies with $M$ from --20 to --18). This is due to the fact
that bright galaxies (compared to galaxies with $M$ from --20 to
\mbox{--18)} have slightly larger sizes. That is why we slightly
increased
the ``fixed galaxy size''.
\subsection{Surface Brightness Evolution}
\begin{figure*}[htbp]
\includegraphics[scale=1.0]{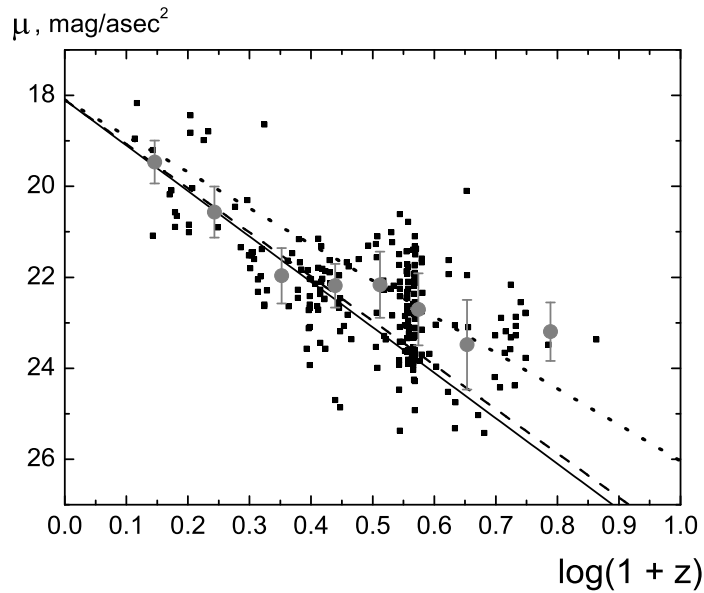}
\caption{Surface brightness evolution as a function of $z$ for the
$C_{i}\_bviz\_mod3\_M20\_22$ sample. The dashed line corresponds to averaging over all points of the sample,
$p = -0.8251$; the dotted line corresponds to averaging over median points, $p=-0.1144$; the solid line shows
the evolution of surface brightness, $p=0$, and the vertical bars show the errors of the median values.}
\label{mu_norm_M-22_-20_fitting:Nabokov_n}
\end{figure*}

The distribution of surface brightness values of the galaxies can
be characterized by the $J(\Theta, z)$ profile for each $z$. In
this paper we analyze the evolution of the surface brightness of
galaxies, i.e., the central value of the  $J(z)_{\Theta=0}$
profile found using  sExtractor and GALFIT codes.

We determine the normalizing constant $J_0$ by averaging the surface-brightness values for galaxies with $z$ from
0 to 0.5. The surface-brightness parameter $n = 4 + p = 4 +$ k $ + e_J$ is given by the following formula:
\begin{equation}
    \label{eq3737:Nabokov_n}
    J_{norm} = \frac{1}{(1 + z)^n},
\end{equation}
where $J_{norm}$ is the normalized surface brightness: $J_{norm}$
=  $J / J_0$ ($J_0$ is the central surface brightness of the given
galaxy) and $p$ is the parameter of evolution. We determine
parameter $n$ by applying the least-squares method both to all
points and to the median values of normalized surface brightness.
In Fig.~\ref{Inorm_M-22_-20_fitting:Nabokov_n} we compare the
curves of the theoretical and observed evolution of surface
brightness.

For better visualization, these curves can be drawn in other axes.
We use formula (\ref{eq3737:Nabokov_n}) to change to surface
brightness $\mu$ :
\begin{eqnarray}
-2.5 \lg(J_{obs}\Delta x^{-2}) = -2.5\lg(J_0 \Delta x^{-2}) + 2.5\lg(1 + z)^n,
\end{eqnarray}
\noindent whence it follows that
\begin{equation}
 \mu_{obs} = \mu_0 + 2.5\,n\,\log(1 + z),
\end{equation}
where $\Delta x$ is the scale factor for converting pixels to
acrseconds. In Fig.~\ref{mu_norm_M-22_-20_fitting:Nabokov_n} we
show the plots with different evolution parameters $p$ in the
($\mu~\mbox{vs.}~\lg(1 + z)$) axes, and $\mu_0 = 17.89$
mag$^{''}$.

\begin{figure}[htbp]\label{zDistrib02:Nabokov_n}
\includegraphics[scale=0.8,bb=12 17 308 239,clip]{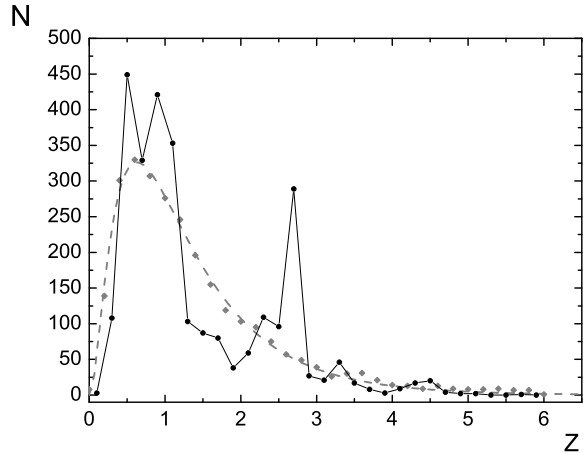}
\caption{Observed (the solid line) and modeled (the dashed line) binned ($\Delta z = 0.2$)
redshift distributions.}
\end{figure}

The results of our analysis of the surface-bright\-ness evolution
of the galaxies of the HUDF field considered imply a
surface-brightness evolution parameter of {\small \mbox{$p=-0.83
\pm 0.1$ $(n=4+p=3.17 \pm 0.1)$}} if computed using all points of
the sample and  {\small \mbox{$p=-0.11 \pm 0.1$ $(n=3.89 \pm
0.1)$}} if computed by using only median points for galaxies in
the interval of absolute magnitudes $M$ \mbox{from --18 to --20.}

\section{IDENTIFICATION OF POSSIBLE SUPERLARGE STRUCTURES IN THE RADIAL DISTRIBUTION OF  HUDF GALAXIES}

The distribution of photometric redshifts of HUDF galaxies in the
z-interval from  0.1 to 6.5 can be searched for superlarge
structures, which show up in the form of fluctuations of the
number of galaxies within the corresponding bins of the
distribution considered.

\subsection{Distribution of Photometric Galaxy Redshifts}

As a model distribution of galaxies to compare with the observed distribution, we use a uniform distribution
of points inside a unit-radius sphere. We randomly assigned to each point an absolute magnitude in accordance
with the Schechter luminosity function:
\begin{equation}
\label{eqSchechter:Nabokov_n}
\phi(L)dl=\phi^*{\left(\frac{L}{L^*}\right)}^\alpha \exp
\left(-\frac{L}{L^*}\right)d\left(-\frac{L}{L^*}\right).
\end {equation}

Our next step was to find the redshifts for modeled data points.
To do this, we had to change from the radial distance unit to
metric distance one. At $k = 0$ (a zero-curvature space) the outer
and inner metric distances coincide, $l(z) = r(z)$. We thus use
the inverse relation $z = l^{-1}(z)$ and formula
(\ref{l-z-m+v-0:Nabokov_n}) to compute the redshift. To allow for
selection due to the limit of telescope sensitivity, we limit the
sample by apparent magnitude---it must not exceed $29^m$.

We fit the resulting model distribution by the following formula:
\begin{equation}\label{eqZdisrtShape:Nabokov_n}
dN = Ax^\alpha\exp\left(-\frac{x}{x_0}\right)^\beta dz,
\end{equation}
where the free parameters   $\alpha$, $\beta$, and $x_0$ are inferred using the least squares method and
$A$ is the normalizing constant.

\begin{table*}[!ht]
\caption{\footnotesize{Results of
computation  of the parameter k of the evolution of angular sizes
of the galaxies for different cosmological models}}
\begin{tabular}{|c|c|c|c|c|}
\hline
    N& Model & \multicolumn{2}{|c|}{Evolution parameter $k$}& $M$(mag.)\\
\hline
    & & based on all points & based on median points&\\
\hline
    I.1 & $ \Omega_m = 0.0, \Omega_V = 1.0$ & $ -0.40 \pm 0.03 $ & $ -0.58 \pm 0.09 $ &-20\,-18 \\
\hline
    I.2 & $ \Omega_m = 1.0, \Omega_V = 0.0$ & $ -0.99 \pm 0.04 $ & $ -1.09 \pm 0.05 $ &-20\,-18 \\
\hline
    I.3 & $ \Omega_m = 0.3, \Omega_V = 0.7$ & $ -0.79 \pm 0.03 $ & $ -0.91 \pm 0.05 $ &-20\,-18 \\
\hline
    I.4 & $ \Omega_m = 0.0, \Omega_V = 0.0$ & $ -0.67 \pm 0.03 $ & $ -0.89 \pm 0.18 $ &-20\,-18\\
\hline
\hline
    II.1 & $ \Omega_m = 0.0, \Omega_V = 1.0$ & $ -0.49 \pm 0.06 $ & $ -0.60 \pm 0.08 $ &-22\,-20\\
\hline
    II.2 & $ \Omega_m = 1.0, \Omega_V = 0.0$ & $ -1.14 \pm 0.10 $ & $ -1.32 \pm 0.12 $ &-22\,-20 \\
\hline
    II.3 & $ \Omega_m = 0.3, \Omega_V = 0.7$ & $ -0.90 \pm 0.08 $ & $ -1.07 \pm 0.06 $ &-22\,-20 \\
\hline
    II.4 & $ \Omega_m = 0.0, \Omega_V = 0.0$ & $ -0.75 \pm 0.08 $ & $ -0.85 \pm 0.12 $ &-22\,-20\\
\hline
\end{tabular}
\label{FWHM_flux_values:Nabokov_n}
\end{table*}

\subsection{Comparison of the Expected and Observed Distributions}

In Fig.~8 we present the observed and modeled (according to
(\ref{eqZdisrtShape:Nabokov_n})) redshift distributions of HUDF
galaxies. The parameters of the modeled distribution are
 $\alpha = 2.84$, \mbox{$\beta = 0.48$}, \mbox{$x_0 = 0.015$.}

\begin{figure}[htbp]\label{ObsDevHUDF02:Nabokov_n}
\includegraphics[scale=0.8,bb=10 15 279 236,clip]{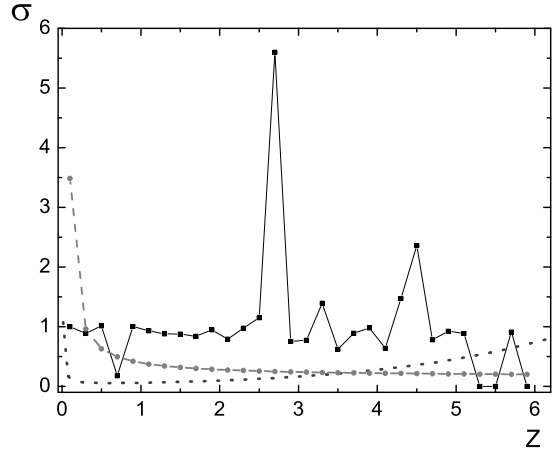}
\caption{Observed (the solid line), theoretical (the dashed line), and Poisson
(the dotted line) deviations of photometric redshifts in  \mbox{$\Delta z =
0.2$} bins.}
\end{figure}

We use the following quantity to measure the deviation of the observed number of galaxies from the theoretically
expected numbers:
\begin{equation}\label{eqDeviations:Nabokov_n}
    \sigma_{obs}=\frac{|N_{obs} - N_{theor}|}{N_{theor}},
\end{equation}
where $N_{theor}$ is the expected  (according
to~(\ref{eqZdisrtShape:Nabokov_n})) number of galaxies in the
interval of redshifts from $z$ to $z+\Delta z$ and $N_{obs}$ is
the number of galaxies observed in the same interval.

The theoretically expected amplitude of fluctuations is characterized by the dispersion of Poisson noise,
$\sigma_{_P}=1/N_{theor}$, and the dispersion associated with correlated structures, which we computed
by the following formula \cite{Peebles:Nabokov_n}:
\begin{equation}\label{eqDeviationsTheor:Nabokov_n}
    \sigma_{theor}^2 = \frac{J_2}{1+z}\,\left(\frac{r_0}{r_{eff}}\right)^\gamma,
\end{equation}
where  $\gamma = 1.8$;
$J_2=72.0/[2^\gamma(3-\gamma)(4-\gamma)(6-\gamma)]$=1.865; $r_0 = 5$ Mpc; $r_{eff} = (3/4\pi r^2 \Delta r
S)^\frac{1}{3}$ is the effective radius corresponding to the
volume of the interval; $\Delta r$ corresponds to the $dz$ layer,
and S is the solid angle of the HUDF field.

In Fig.~9 we present the plot of observed deviations
$\sigma_{obs}$ and theoretically expected deviations $\sigma_{p},
\sigma_{theor}$ from uniform distribution of galaxies.

\section{DISCUSSION OF THE RESULTS AND THE MAIN CONCLUSIONS}

Our analysis of the classical cosmological test $\Theta(z)$ shows that the choice of cosmological models has a
strong effect on the parameter of the evolution of linear sizes of galaxies.

It is evident from Table~\ref{FWHM_flux_values:Nabokov_n} that the
parameter of the evolution of linear sizes for galaxies with
absolute magnitudes in interval from --20 to --18 varies from
k$=-0.40 \pm 0.03$ for the model with \mbox{$\Omega_m = 0.0$}\,
and\, $\Omega_V = 1.0$ to k$=-1.09 \pm 0.06$ for the model with
\mbox{$\Omega_m = 1.0$} and $\Omega_V = 0.0$. This parameter
varies from $-0.49 \pm 0.06$ to $-1.32 \pm 0.12$ for galaxies with
luminosities in the interval from --22 to --20. The inferred
values of parameter k agree with the results of Bowens et al.
\cite{Bouwens:Nabokov_n} for other samples of galaxies from HDF-S,
HDF-N, GOODS, and HUDF. The above authors used a cosmological
model with parameters $\Omega_m = 0.3\, , \, \Omega_V = 0.7 $. For
galaxies with luminosities in the interval of absolute magnitudes
$M$ from -22.38 to \mbox{-21.07} the parameter of galaxy size
evolution was found to be k$=-1.05 \pm 0.21$. For our sample
parameter k$ = -1.07 \pm 0.06$ in case of $\Omega_m = 0.3, \,
\Omega_V = 0.7$. Note that the $\Theta(z)$ diagram may become an
efficient cosmological test when a reliable model of the evolution
of galaxy sizes is developed.

The surface brightness evolution parameter  \linebreak $p = -0.11
\pm 0.1~ (n = 3.89 \pm 0.1)$ inferred from the median points for
galaxies in the absolute magnitude ($M$) interval from -18 to -20
requires further analysis. This is due to the fact that the
K-correction to the surface brightness includes a combination of
the K-correction to the flux and the K-correction to the angular
size.

An analysis of the distribution of HUDF galaxies reveals strong
deviations of the observed number of galaxies from the number of
galaxies expected for a uniform distribution. The observed
irregularities correspond to a scale length of about  2000~Mpc.
This may be due both to real superlarge structures and to hidden
selection effects that show up in finding the photometric
redshifts. This problem requires further analysis.

\begin{acknowledgments}
We are grateful to N.~Lovyagin for sharing the results of his computations of the radial distributions of galaxies
in simulated catalogs. This work was supported in part by the Scientific School and Rosobrazovanie foundations.
\end{acknowledgments}

\end{document}